\documentclass[conference]{IEEEtran}
\usepackage{cite}
\usepackage{amsmath,amssymb,amsfonts}
\usepackage{algorithmic}
\usepackage{graphicx}
\usepackage[caption=false]{subfig}
\usepackage{textcomp}
\usepackage{xcolor}
\def\BibTeX{{\rm B\kern-.05em{\sc i\kern-.025em b}\kern-.08em
    T\kern-.1667em\lower.7ex\hbox{E}\kern-.125emX}}

\usepackage{geometry}
\begin{document}

\title{Orthogonal Circular Polarized Transmitter and Receiver Antennas for Mitigation of Mutual Coupling in Monostatic Radars\\
}

\author{\IEEEauthorblockN{Shobha Sundar Ram}
\IEEEauthorblockA{\textit{Department of ECE} \\
\textit{Indraprastha Institute of Information Technology}\\
New Delhi, 110020 India \\
shobha@iiitd.ac.in}
\and
\IEEEauthorblockN{Akanksha Sneh}
\IEEEauthorblockA{\textit{Department of ECE} \\
\textit{Indraprastha Institute of Information Technology}\\
New Delhi, 110020 India \\
akankshas@iiitd.ac.in}

}

\maketitle

\begin{abstract}
Through-wall radar systems require compact, wideband and high gain antennas for detecting targets. Building walls introduce considerable attenuation on the radar signals. When the transmitted power is raised to compensate the through-wall attenuation, the direct coupling between the transmitter and receiver can saturate the receiver because of which weaker reflections off the target may remain undetected.
In this paper, we propose using transmitter and receiver antennas of orthogonal circular polarization to reduce the direct coupling between the transmitter and receiver while retaining the first bounce off the target. In our paper, we demonstrate that the quadrafilar helical antenna (QHA) is a good candidate for this operation since it is characterized by a small size, wide frequency band of operation, high gain and low axial ratio over a wide field of view. We compare the reduced mutual coupling between the transmitter and receiver elements for the oppositely polarized QHA antennas with other commonly used through-wall radar antennas such as the Vivaldi and horn antennas. The system is tested in through-wall conditions. 

\end{abstract}


\section{Introduction}

\label{sec:Intro}
Through-the-wall radars (TWR) are being developed for law enforcement, security and bio-medical applications \cite{b1, b2}. These radars may be used to detect both static and dynamic targets - mostly humans - as well as to estimate building layouts. Hence, they are usually characterized by wide bandwidths (usually greater than 1GHz) for obtaining fine down-range resolutions and must have high enough transmit power to overcome the attenuation introduced by two-way propagation through different walls. The attenuation of the radar signal through walls increases with the carrier frequency. Most researchers have found the frequency range of 1-4GHz to be the most optimal range for through-wall applications. The antennas of through-wall systems must meet the following specifications: they must be compact and of low profile to enable a portable radar system; they should be wideband and must be of high gain. Most wideband through-wall radar systems either operate in the stepped frequency or frequency-modulated continuous wave mode rather than the impulse mode \cite{b3}. This is to enable Doppler-based detection of indoor movers \cite{b4, b5}.  In this situation, the transmitter is not switched off while the receiver is operational. Circulators are not widely used in this application due to the weak isolation between the transmitter and receiver channels. Instead, the transmitter and receivers are configured with separate antennas. The maximum transmitted power across a wideband system is specified by FCC regulations. However, the radar may frequently be operated at a sub-optimal transmitted power due to the strong mutual coupling between the transmitter and receiver antennas. When the transmitted power exceeds a certain threshold, the receiver channel gets saturated resulting in the suppression of the weaker returns from the targets. Hence, the mutual coupling between the two antennas becomes a critical parameter that determines the maximum distance that the radar can see "through" a wall.

Several different types of antennas have been examined for through-wall radar applications. Printed microstrip antenna elements are lightweight, compact and of low profile with the disadvantage of narrow bandwidth \cite{b4, b6}. Many  methods have been researched to increase the bandwidth such as the introduction of slots within the patch and stacking of multiple dielectric layers \cite{b7, b8}. When these antennas are configured as linear arrays, the gain can be enhanced. Thus, low profile linearly polarized TWR arrays can be designed with microstrip antennas with bandwidth enhancement techniques. However, they usually have high mutual coupling between the transmitter and receiver. Horn and log periodic antennas are other examples of ultra wideband antennas that have been used in through-wall sensing, but they have the disadvantage of bulky size \cite{b9, b3, b10, b11, b12}. Thus, they are not very suitable for portable systems. Of late, planar Vivaldi antennas have been identified as excellent candidates for TWR systems because of the simplicity of design, compact size, wide bandwidth and high gain characteristics \cite{b13, b14, b15}. However, there is still the problem of mutual coupling between the transmitter and receiver elements. 
In this paper, we propose to tackle the problem of mutual coupling between the receiver and co-located transmitter antenna by using two orthogonal circularly polarized antennas. When the transmitter and receiver are of orthogonal circular polarization, the direct coupling between them is reduced. On the other hand, the primary reflections off the target undergo sense reversal. As a result, the signal from the target is picked up without polarization mismatch loss. There are many candidates of circularly polarized antennas. In this work, we propose the use of quadrifilar helix antennas (QHA) for TWR applications. They have the advantage of wide bandwidth, high gain and a very compact size. Preliminary results indicate that these orthogonally polarized antennas are successful in mitigating mutual coupling in indoor line-of-sight conditions \cite{b16}. In this work, the antennas are mounted on a platform and the whole radar system is tested with target and without target in through-wall conditions. Their performances are bench-marked against other commonly used through-wall radar antennas.

The paper is organized as follows. In the following section, we discuss the proposed solution of reduced mutual coupling using orthogonal circularly polarized antennas. In section 3, we discuss the design of QHA antennas and their feed network. We present the experimental results in section 4 followed by the conclusion.

\section{Proposed Method for Mitigating Mutual Coupling}
\begin{figure}[htbp]
\vspace{3mm}
            \centering
\includegraphics[width = 0.4\textwidth, height = 2in]{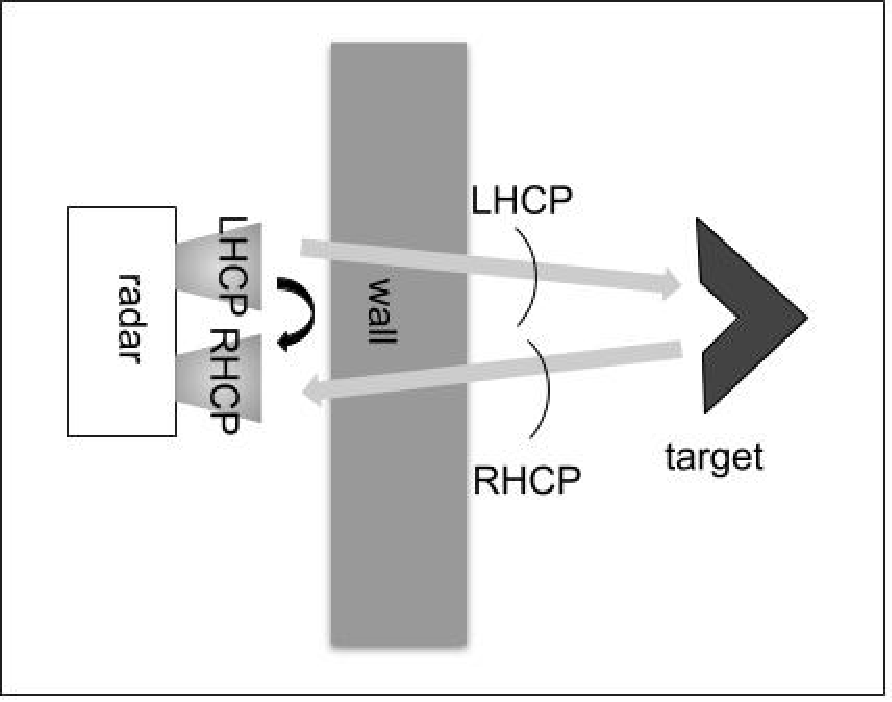}
    \caption{Reduced mutual coupling due to orthogonal circular polarized antennas in through-wall radar}
    \label{fig:System_Fig}
\end{figure}

Fig.\ref{fig:System_Fig} shows a quasi-monostatic radar deployed in a through-wall scenario. The radar consists of separate transmitter and receiver antennas to facilitate a 100\% transmission duty cycle. To the best of our knowledge, current through-wall radar systems have deployed linearly polarized antennas. As a result, the following signals will superpose at the receiver - the direct coupling from the transmitter; reflections from the front face of the wall and from multiple reflections within the wall; scatterings from the target; other multipath signals and interference from the environment (not shown here). Usually, the reflections off the target are weak due to the two-way attenuation introduced by the walls. If the transmitted power is increased to improve the maximum detectable range, then the direct coupling between the transmitter and receiver proportionately increases. The maximum transmitted signal is, thereby, limited by the minimum signal required to saturate the receiver. We propose to mitigate this problem by introducing orthogonal circular polarized antennas at the transmitter and receiver. Provided the antennas have a low axial ratio over a wide field-of-view, the mutual coupling between the antennas will be reduced. The first (and other odd) reflections off the target will undergo polarization reversal. Therefore, the receiver will be capable of receiving the scatterings from the target. The proposed solution may however not be very effective in reducing the direct scatterings from the wall. These are typically removed using the process of background subtraction. In background subtraction, measurements are first made in the absence of a target. These signals are subsequently coherently subtracted from the signals received after the introduction of a target.

\section{Choice of antennas for through-wall radars}
Many different types of circularly polarized antennas have been realized in literature. Some of the most compact designs are printed microstrip antennas specifically designed for circular polarization. However, these antennas are characterized by a low field-of-view around the broad side over which they have a low axial ratio. Therefore, when these antennas are placed side by side in a monostatic radar system, they appear linearly polarized along the end-fire direction and hence, demonstrate high mutual coupling. Alternatively, helical antennas can be considered. They are characterized by high gain and high bandwidths. They also have a high field-of-view over which the axial ratio is low. 
The drawback of the helical antenna is its large size, which makes it unsuitable for portable radar applications.
\subsection{Quadrifilar helix antenna}
\begin{figure*}[htbp]
\vspace{0.4in}
\centering
\subfloat{
\includegraphics[width=0.25\textwidth ,height = 1.5in]{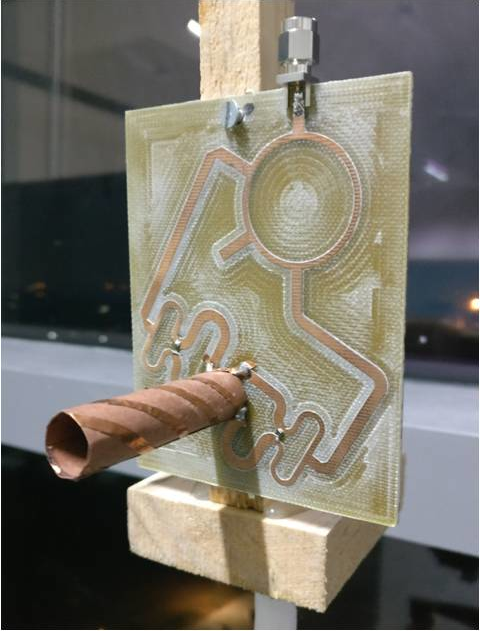}}
\hfil
\subfloat{
\includegraphics[width=0.25\textwidth ,height = 1.5in]{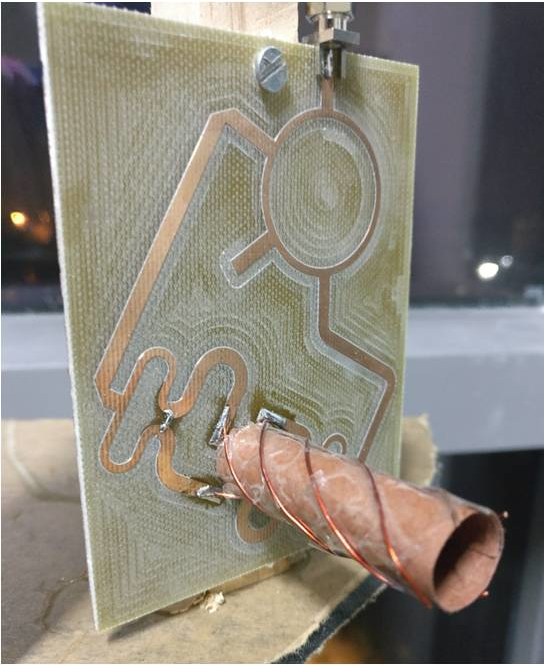}}
\hfil
\subfloat{
\includegraphics[width=0.25\textwidth ,height = 1.5in]{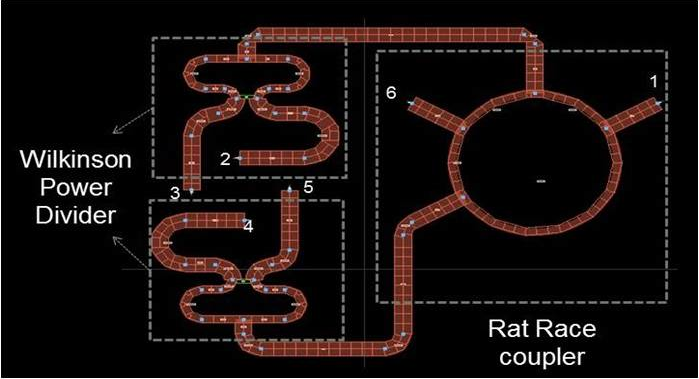}}
\caption{(a) Printed quadrafilar helical antenna (b) Wire wound quadrafilar helical antenna (c) Quadrature feed network}
\label{fig:QHA}
\end{figure*}

The quadrifilar helix antenna (QHA) is a compact alternative that provides circular polarization with very low sidelobe levels \cite{b17,b18}. This antenna has 4 parallel helical windings which are excited with a progressive phase excitation of $90^{\circ}$. The extra windings tighten up the radiation pattern and reduce the sidelobe level. The clockwise phase progression induces an end-fire radiation pattern. The polarization of the antenna depends upon the sense of the helix winding, either right-handed or left-handed. The presence of the ground plane reverses the polarization of the antenna. Another version of QHA is the printed quadrifilar helix antenna (PQHA) where the arms of the antenna are printed on the dielectric and then folded in the shape of a cylinder. In this work, both printed and wire QHA have been designed and fabricated for a resonant frequency of 3.4GHz. They are shown in Fig.\ref{fig:QHA}a and Fig.\ref{fig:QHA}b. The design parameters for both the QHA are given in Table.\ref{tab:QHA}. The performances of both QHAs were found to be comparable.
\begin{table}[b!]
    \centering
    \begin{tabular}{c|c}
    \hline \hline
    Design parameters & Values \\
    \hline
         height of QHA & $4.5 cm$ \\  
         radius OF QHA & $0.5 cm$ \\
         Number of turns & 1 \\
         Feed circuit & $7.5 \times 9.5 cm$\\
        \hline \hline
    \end{tabular}
    \caption{Design parameters of the quadrifilar helix antenna}
    \label{tab:QHA}
\end{table}
The QHA height is reduced to less than half of that of a helical antenna resonating at the same frequency. 
The feed network required for providing the $0^{\circ}, 90^{\circ}, 180^{\circ}$ and $270^{\circ}$ degrees phase shifts is shown in Fig.\ref{fig:QHA}c. Here, port 1 is the input port. The feed network has a rat race coupler and a Wilkinson power divider to provide the required shifts to the four output ports (2,3,4,5). 

\subsection{Comparison of Antennas for TWR Purposes}
\begin{figure*}[htbp]
\vspace{4mm}
\centering
\subfloat{
\label{fig:Vivaldi}
\includegraphics[width=0.25\textwidth ,height = 1.5in]{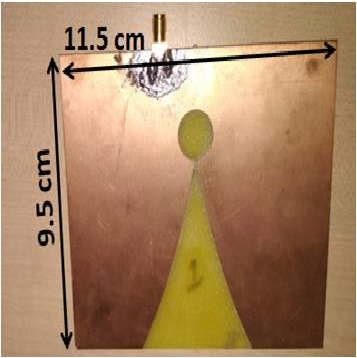}}
\hfil
\subfloat{\label{fig:AP_Vivaldi}
\includegraphics[width=0.25\textwidth ,height = 1.5in]{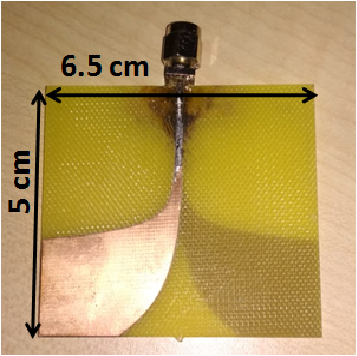}}
\hfil
\subfloat{
\label{fig:Horn}
\includegraphics[width=0.25\textwidth ,height = 1.5in]{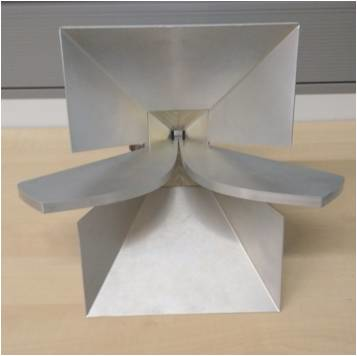}}
\hfil
\caption{Commonly used through-wall radar antennas (a) Linear polarized Vivaldi (b) Linearly polarized antipodal Vivaldi (c) Linearly polarized broadband horn}
\label{fig:OtherAntennas}
\end{figure*}

We compare the characteristics of the proposed QHA with other commonly used antennas for TWR systems - a regular Vivaldi, an antipodal Vivaldi and a Rohde and Schwarz HF907 horn antenna. All three antennas are shown in Fig.\ref{fig:OtherAntennas}. The two Vivaldi antennas were fabricated in-house. All of these antennas are linearly polarized, broadband, and can be operated over the same band of frequencies as the QHA. The size, gain, bandwidth, and axial ratio of the antennas are compared in Table.\ref{tab:AntennaCompare}. The gain and axial ratio are reported at 3.4GHz. Finally, their mutual coupling is measured for the case when two similar types of antennas are configured as transmitter and receiver. The antennas are placed 15 cm apart and the measurements are made in an anechoic chamber in the absence of a target. 
Of the four antennas, the horn antenna offers the highest gain, the largest bandwidth, and the lowest coupling between the transmitter and receiver. However, the antenna is large and heavy. Both the Vivaldi antennas are compact in size and of high bandwidth. The antipodal Vivaldi is smaller than the regular Vivaldi antenna. However, in both cases, the isolation between the transmitter and receiver elements is poor since the antennas are planar and of similar polarization. The QHA is not planar but it is still small in size. Its bandwidth is satisfactory for most TWR applications. Most importantly, when two QHAs of orthogonal polarization (OP) are used as transmitter and receiver respectively, their mutual coupling gets mitigated by 8dB at 3.4GHz when compared to the case when antennas of similar polarization (SP) are used. 
\begin{table*}[htb]
    \centering
    \begin{tabular}{p{2.5cm} p{2cm} p{2cm} p{2cm} p{2.5cm} p{2cm}}
    \hline \hline
        Antenna & Gain & Axial Ratio & Bandwidth & Size & Tx-Rx Coupling \\
        & ($dBi$) & ($dB$) & ($GHz$) & ($cm$) & ($dB$)\\
        \hline \hline
         QHA & 4 & 2 & 1.5 & $9.5 \times 7.5 \times 4.5 $ & -31 (OP) \\
          & & & & & -23 (SP) \\
          \hline
         Vivaldi & 6 & 24 & 3 & $11.5 \times 9.5$ & -23 (SP) \\
         Antipodal Vivaldi & 2 & 17 & 3 & $6.5 \times 5$ & -19 (SP) \\
         Horn & 9 & 20 & 17 & $31 \times 28$ & -37 (SP) \\
         \hline \hline
    \end{tabular}
    \caption{Comparison of antennas for through-wall radar applications. The last column indicates the mutual coupling between transmitter (Tx) and receiver (Rx) antennas of similar (SP) and opposite polarization (OP).}
    \label{tab:AntennaCompare}
\end{table*}

\section{Experimental Results}
\begin{figure*}[htbp]
\vspace{0.4in}
\centering
\subfloat{
\includegraphics[width=0.4\textwidth, height = 0.2\textheight]{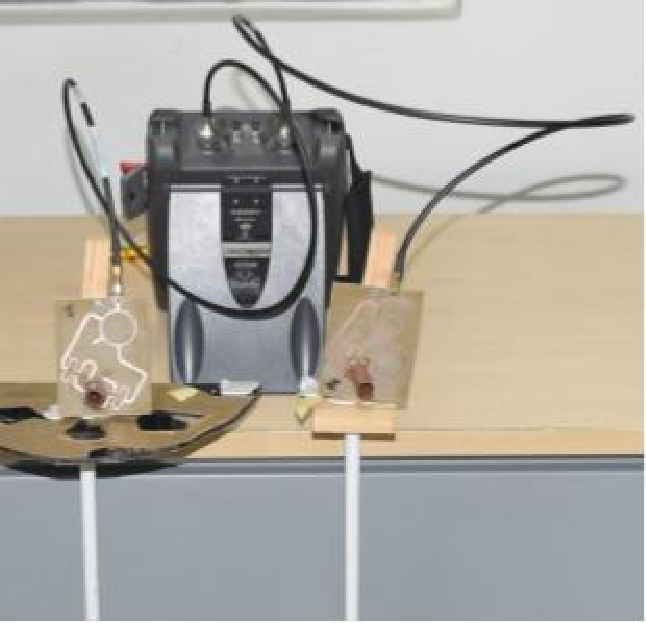}}
\hfil
\subfloat{
\includegraphics[width=0.4\textwidth, height = 0.2\textheight]{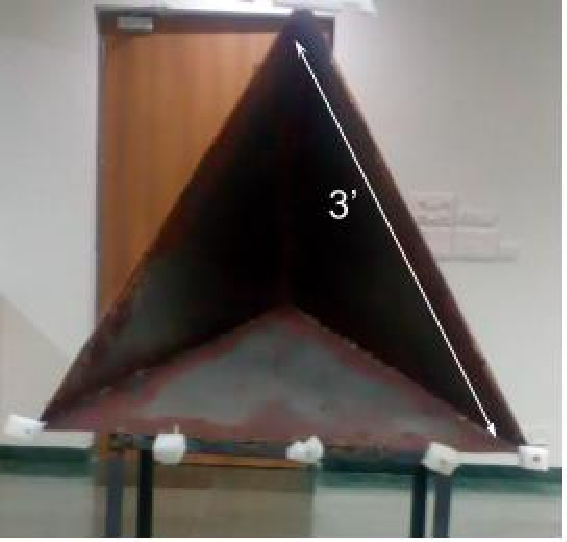}}
\hfil
\caption{Wideband monostatic radar using two orthogonal circular polarized quadrafilar helical antennas and a vector network analyzer (b) Trihedral corner reflector used as radar target}
\label{fig:SetUp}
\end{figure*}

\begin{figure*}[htbp]
    \centering
    \includegraphics[scale = 0.3]{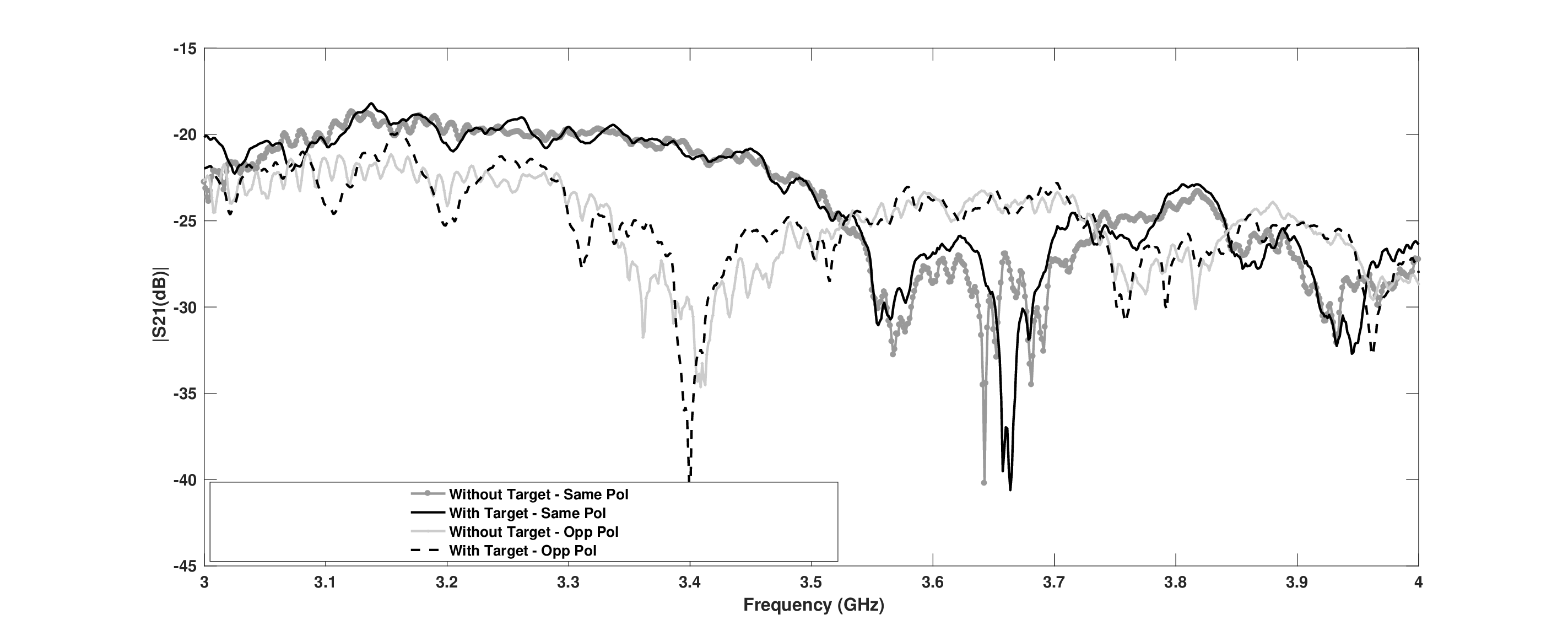}
    \caption{$S_{21}$ measurements with VNA when transmitter and receiver have (1) the same sense of polarization (SP) (2) opposite senses of polarization (OP)}
    \label{fig:S21wwoTargets}
\end{figure*}

In this section, we test the performance of the proposed antennas in a monostatic radar configuration. The antennas are connected to two ports of a Field Fox vector network analyzer (VNA), N9926A, on which broadband frequency domain $S_{21}$ measurements are made as shown in Fig.\ref{fig:SetUp}. 
We consider a trihedral reflector, shown in Fig.\ref{fig:SetUp}b, as the target. The transmitted power is $+3dBm$, the maximum power that can be generated in the VNA. When two antennas of similar polarization (horn, Vivaldi) are connected to the two ports of the VNA, the receiver channel is saturated even in free space conditions. Next, two QHA are connected to the two ports of the VNA. The radar is turned towards an indoor two-layer brick wall of approximately 18 inch thickness. The radar is placed approximately 15 cm before the wall while the target is placed on the other side of the wall at a distance of 2m from the radar. The experimental setup thus resembles Fig.\ref{fig:System_Fig}. We make the measurements for the following cases -
\begin{enumerate}
    \item We consider similarly polarized - left hand circularly polarized - transmitter and receiver QHAs placed along side each other in the absence of target. We measure the $S_{21}$ over a bandwidth of 1GHz about the resonant frequency of 3.4GHz. 
    \item We introduce the trihedral reflector to the above scenario and repeat the measurements.
    \item Next we consider orthogonally polarized antennas as transmitter and receiver in the same scenario in the absence of a target.
    \item Finally, the measurements are repeated with the orthogonally polarized antennas with the presence of the trihedral target.
\end{enumerate}
When both antennas are of the same polarization and at a distance of 15cm from each, there is strong mutual coupling between them as shown in Fig.\ref{fig:S21wwoTargets}. The receiver is, in this case, saturated with the direct signal from the transmitter. When the target is introduced, there is a negligible change in the $S_{21}$ across the bandwidth. As a result, the target is not discernible even at close ranges from the radar. Note that the direct coupling cannot be removed through background subtraction operations since the receiver is saturated. If we reduced the transmitted power, we would no longer saturate the receiver and the mutual coupling may be mitigated through some form of background subtraction. However, the reduced transmitted power would lower the maximum detectable range. 
When we consider two antennas of opposite polarization, we immediately observe a significant reduction in the mutual coupling in the figure when compared to the case of the same polarization. Now, we introduce the target to this radar configuration. The target returns can now be clearly discerned by the receiver as there is a jump in the peak $S_{21}$. The target's range may be estimated by background subtraction in the frequency domain followed by the Fourier transform of the subtracted signal. 
\begin{table*}[htbp]
    \centering
    \begin{tabular}{ccc}
      \hline \hline
      TX - RX configuration & Scenario & Peak difference in $S_{21}$ \\
      \hline \hline
        LHCP - LHCP & Presence of target & 2 dB  \\
        LHCP - LHCP & Absence of target & \\
        \hline
        LHCP - LHCP & Absence of target & 9 dB  \\
        LHCP - RHCP & Absence of target & \\
        \hline
        LHCP - RHCP & Presence of target & 6 dB  \\
        LHCP - RHCP & Absence of target & \\
        \hline \hline
    \end{tabular}
    \caption{Peak $S_{21}$ across 3 to 4GHz for different transmitter (TX) - receiver (RX) antenna configurations in the presence/absence of trihedral target}
    \label{tab:ExpResults}
\end{table*}
In other words, the configuration of the orthogonally polarized antennas, allows us to use a high transmit power on the radar. This is particularly useful when the radar is deployed in a through-wall scenario since walls introduce considerable two-way attenuation of the radar signal.
For each of the above cases, measurements are made for different distances between the transmitter and receiver antennas -15cm, 20cm and 25cm. Likewise, measurements were made for different distances of the trihedral target from the radar - 0.5m to 3m with measurements at every 0.5m.  We report the average results from all of these measurements in Table.\ref{tab:ExpResults}. The results indicate that a target may not be detected if the receiver is saturated with a strong transmitted signal. However, when orthogonally polarized antennas are used for the transmitter and receiver, there is a significant reduction in the mutual coupling which allows the detection of a target. 

\section{Conclusion}
Mutual coupling between transmitter and receiver antennas in a monostatic radar configuration limits the maximum power that can be transmitted. Reduced coupling is achieved by using orthogonal circularly polarized antennas. Odd bounces from the scatterer cause inversion of polarization, resulting in enhanced target detection.
Quadrifilar helix antenna is a good candidate for through-wall radars due to its high gain, bandwidth, compact size and low axial ratio over a high field-of-view.

\section*{Acknowledgement}
This work is sponsored by the 5IOA036 FA23861610004 grant by the Air Force Office of Scientific Research (AFOSR), AOARD.

\end{document}